\documentclass[reprint,groupedaddress,amsmath,amssymb,aip,jap,floatfix]{revtex4-1}
\usepackage[version=3]{mhchem}
\usepackage{siunitx,natmove,graphicx}

\begin{document}

\title{Interfacial Mechanical Behaviors in Carbon Nanotube Assemblies}
\author{Xiaohua Zhang}
\email{zhangxhcm@gmail.com}
\affiliation{Suzhou Institute of Nano-Tech and Nano-Bionics, Chinese Academy of Sciences, Suzhou 215123, China}
\affiliation{Department of Engineering Mechanics, School of Civil Engineering, Southeast University, Nanjing 210096, China}

\begin{abstract}
Interface widely exists in carbon nanotube (CNT) assembly materials, taking place at different length scales. It
determines severely the mechanical properties of these assembly materials. Here I assess the mechanical properties of
individual CNTs and CNT bundles, the inter-layer or inter-shell mechanics in multi-walled CNTs, the shear properties
between adjacent CNTs, and the assembly-dependent mechanical and multifunctional properties of macroscopic CNT fibers
and films.
%\keywords{carbon nanotube, assembly materials, mechanics, interface}
\end{abstract}

\maketitle

\section{Introduction}

Carbon nanotubes (CNTs) are hollow cylinders consisting of single or multiple sheets of graphite (graphene) wrapped into
a cylinder. After the first observation in 1991 by Iijima \cite{iijima.s_1991}, extensive work has been carried out to
CNTs towards their broad mechanical, electronic, thermal, and optical applications \cite{baughman.rh_2002,
devolder.mfl_2013}. The fundamental mechanical properties, such as their stiffness, strength, and deformability, have
been investigated with extensive theoretical and experimental researches, and been well reviewed, in the past two
decades \cite{qian.d_2002, bernholc.j_2002, ruoff.rs_2003, yu.mf_2004, kis.a_2008}. Nowadays, for industrial and
engineering applications, CNTs have been assembled into macroscopic materials such as fibers, films, forests, and gels
\cite{behabtu.n_2008, liu.lq_2011, zhang.xh_2012, lu.wb_2012, miao.mh_2013, zhang.xh_2016, di.jt_20161}. Due to the
assembly feature of these materials, the interfacial structure between CNTs always plays a crucial role in determining
the tensile or compressive behavior, dynamic response, and coupling phenomenon between multiple physical properties that
take place at the interfaces. Therefore, the utilization efficiency of the mechanical property from individual CNTs to
their assembly is a sophisticated function of the assembly structure and types of interfacial interactions.

Here I will review the mechanics of individual CNTs, discuss the sliding friction between CNTs, and report our recent
progresses on the fabrication technique and mechanical properties of macroscopic CNT assembly materials. The general
features of the assembly design and interface design will be illustrated from the perspective of interface engineering.
This paper is organized as follows. Section \ref{sec.CNT} contains a brief description of the mechanical properties of
individual CNTs, in terms of modulus, strength, compressibility, and deformability. Section \ref{sec.interaction}
describes the inter-layer (inter-shell) sliding of multi-walled CNTs (MWCNTs) and intertube friction between adjacent
CNTs, and the strategies to enhance the interfacial frictions. Section \ref{sec.assembly} discusses the recent
investigations on the interfacial mechanics in macroscopic CNT assembly materials, especially CNT fibers and films where
the CNTs are highly aligned or highly entangled. Finally, the paper concludes with a brief summary and gives an outlook
on future developments in the field.

\section{Mechanical Properties of Carbon Nanotubes}
\label{sec.CNT}

\subsection{Young's modulus and tensile strength}

The basic mechanical properties of CNT are strongly related to the basic properties of a graphene sheet. In these
materials, carbon atoms are covalently bonded with three nearest carbon atoms in a hexagonal lattice by forming three
$\sigma$ bonds in the sheet plane and one delocalized $\pi$-bond over the basal plane. As the sp$^2$ $\sigma$ bond is
shorter and stronger even than the sp$^3$ bond in diamond, the axial elastic modulus of CNT and graphene is super high,
as expected to be up to several TPa \cite{ruoff.rs_2003, yu.mf_2004}. The first experimental study on the modulus was
based on the analysis of thermal vibration of MWCNTs, which showed a wide range of 0.4--4.15 TPa \cite{treacy.mmj_1996}.
Later, the modulus was also estimated by the compressive response using micro Raman spectroscopy and was 2.8--3.6 TPa
for single-walled CNTs (SWCNTs) and 1.7--2.4 GPa for MWCNTs \cite{lourie.o_19982}. Direct tensile tests showed the
modulus ranged from 270 to 950 GPa for individual MWCNTs \cite{yu.mf_2000}, and from 320 to 1470 GPa when several SWCNTs
bundled together as a CNT rope \cite{yu.mf_20001}. Besides the experimental evaluations, theoretical estimations also
show the similar results. For example, By using an empirical Keating Hamiltonian with parameters determined from first
principles, a modulus ranging from 1.5 to 5.0 TPa was estimated \cite{overney.g_1993}. A molecular dynamics (MD)
approach showed a modulus of $\sim$1 TPa and a shear modulus of $\sim$0.5 TPa and predicted that chirality, radius and
number of walls have little effect on the value of Young's modulus \cite{lu.jp_1996}.

\begin{figure}[t!]
\centering
\includegraphics[width=0.45\textwidth]{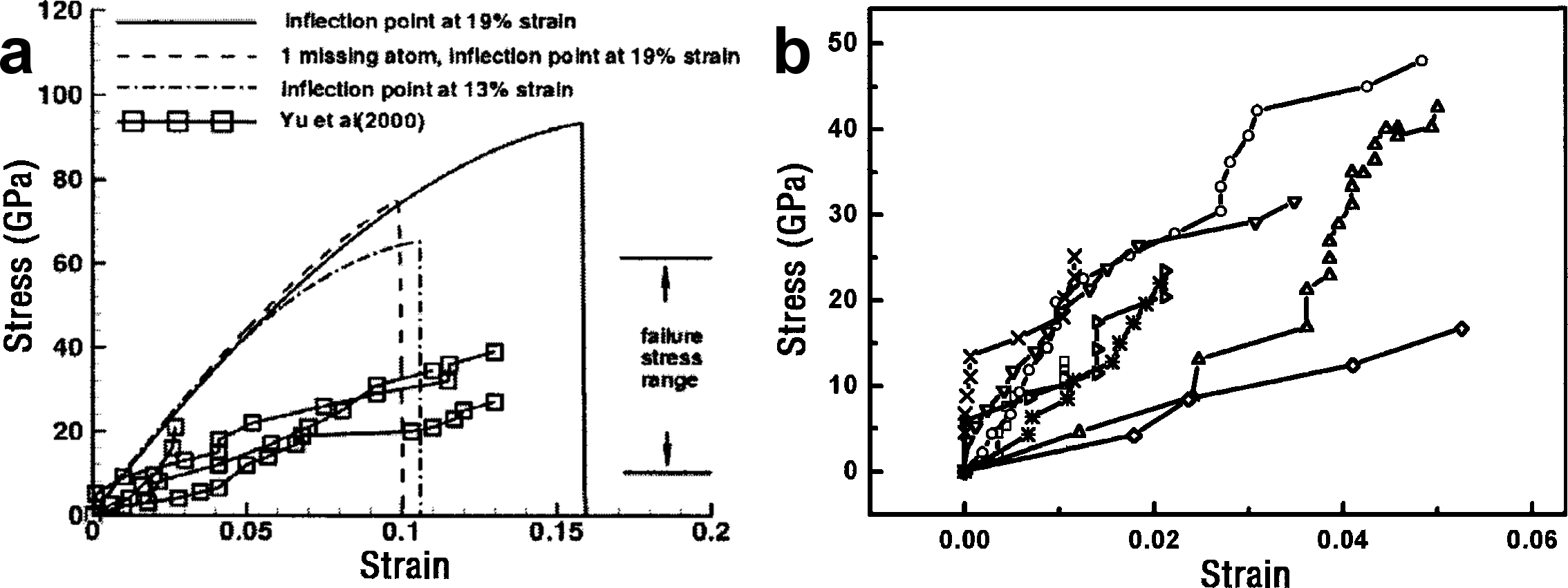}
\caption{(a) A strong dependence on the inflection point in the inter-atomic potential can be seen from the stress-stain
curves for a (20,0) CNT, and a missing atom has also a significant effect on strength, reducing it by about 25\%
\cite{belytschko.t_2002}. Experiment measurement on MWCNTs are provided for comparison \cite{yu.mf_2000}. (b) Eight
stress-strain curves obtained from the tensile-loading experiments on SWCNT ropes \cite{yu.mf_20001}.}
\label{fig.CNTstrength}
\end{figure}

Different from the modulus estimation, measuring the tensile strength of CNTs is a very challenging task. Compared with
experiment, it is easier to compute the strength by considering the effects of defects \cite{belytschko.t_2002,
dumitrica.t_2003}, loading rate, temperature \cite{wei.cy_20031}, and number of walls \cite{liew.km_2004}. Typically,
the fraction of CNT involves bond breakage and/or rotation, usually resulting in the formation of specific types of
dislocation, such as the pentagon-heptagon (5-7), (5-7-7-5), (5-7-5-8-5) defects \cite{bernholc.j_1998}. MD simulations
showed that CNT under tension behaves as a brittle material at high strain (15\%) and low temperature (1300 K), or as a
ductile material at low strain (3\%) and high temperature (3000 K) \cite{bernholc.j_1998, nardelli.mb_1998,
nardelli.mb_19981}. Such tensile behaviors depend on the evolution of the dislocation. For example, the (5-7-7-5)
dislocation can either separate into two (5-7) pairs to result in a ductile transformation or involve into a crack to
result in a brittle fracture. An alternative pathway for the fracture of CNT was also proposed by the direct
bond-breaking through the formation of a series of virtual defects at high tensions \cite{dumitrica.t_2003}. These
studies were usually based on the consideration of the formation energy for the dislocations. However, another MD study
showed that the fracture behavior is almost independent of the separation energy and to depend primarily on the
inflection point in the inter-atomic potential \cite{belytschko.t_2002}. The fracture strength should be moderately
dependent on tube chirality, and the strength and fracture strain were estimated to be 93.5--112 GPa and 15.8--18.7\%,
respectively. Figure \ref{fig.CNTstrength}a shows two different stress-strain curves for a (20,0) CNT, where the
modified Morse potentials with inter-atomic force peaks at 19\% and 13\% strain were used. To show the effect of
structural defect, a result for a missing atom is provided where the strength was reduced by about 25\%.

The tensile strength and the fracture of CNT were also measured directly with experiment \cite{yu.mf_2000, yu.mf_20001}.
The measurement of 19 MWCNTs showed a strength of 11--63 GPa and a strain at break up to 12\% \cite{yu.mf_2000}. The
failure of MWCNTs was described in term of the ``sword-in-sheath'' type fracture, where the outermost shell broke first
followed by the pull out of the rest of the shells from the outermost shell. When SWCNTs were bundled to be a rope, the
tensile strength was measured to be 13--52 GPa, based on 15 measurements, and the strain at break was up to 5.3\%, see
Figure \ref{fig.CNTstrength}b \cite{yu.mf_20001}. The stretchability of CNT was also experimentally obtained by
deforming freely suspended SWCNT ropes, where the maximum strain was up to 5.8$\pm$0.9\% \cite{walters.da_1999}. By
measuring the stress-induced fragmentation of MWCNTs in a polymer matrix, the tube strength was up to 55 GPa
\cite{wagner.hd_1998}. Another pulling and bending tests on individual CNTs in-situ in a transition electron microscope
(TEM) showed a tensile strength of $\sim$150 GPa, suggesting that the strength is a large fraction of the elastic
modulus \cite{demczyk.bg_2002}.

\subsection{Compressibility and deformability}

\begin{figure}[t!]
\centering
\includegraphics[width=0.35\textwidth]{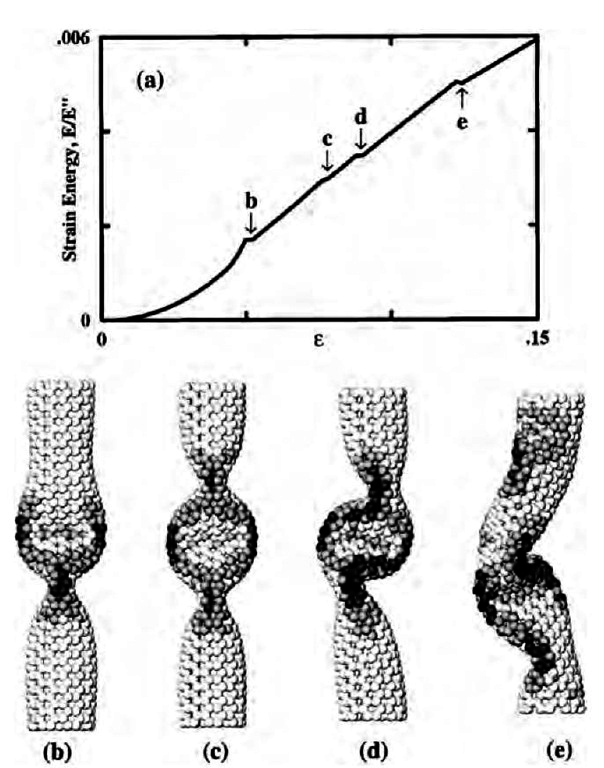}
\caption{The strain energy (a) displays four singularities corresponding to shape changes (b-e) with increasing the
axial compression \cite{yakobson.bi_1996}.}
\label{fig.compressing}
\end{figure}

Different from the high rigidity and high strength along the axial direction, CNT is relatively compressible and
deformable in the transverse direction. When individual SWCNTs are bundled together, significant deformation could occur
by van der Waals (vdW) interaction between the adjacent tubes \cite{tersoff.j_1994}. Fully collapsed MWCNTs were first
observed with TEM \cite{chopra.ng_1995}. Large diameter CNTs could also form the partial and full collapse on substrate
\cite{hertel.t_1998, hertel.t_19981, yu.mf_2001}. The structural deformation can be induced by applying pressure.
Various experiments on SWCNT bundles have shown clear evidence of structural deformation or transition
\cite{chesnokov.sa_1999, venkateswaran.ud_1999, peters.mj_2000, gaal.r_2000, tang.j_2000, lopez.mj_2001}. MD simulations
revealed that the compressibility and deformability are dependent on the tube diameter and the number of walls
\cite{lordi.v_1998}. By applying a certain high pressure, there could be a circular-to-elliptical shape transition of
the tube's cross section for either individual CNTs or their bundles \cite{elliott.ja_2004, zhang.xh_2004,
zhang.xh_20041, sun.dy_2004, ye.x_2005, gadagkar.v_2006, chang.tc_2008, zhang.xh_2010}. These structural deformations
are all reversible upon unloading the pressure. In a different way, a very large force exerted on CNT could still
produce reversible and elastic deformation, and that radial mechanical forces might not be capable of cutting a CNT
\cite{lordi.v_1998}. By examining the ballistic impact and bouncing-back processes, large diameter CNTs could withstand
high bullet speeds \cite{mylvaganam.k_2007}.

\begin{figure}[t!]
\centering
\includegraphics[width=0.40\textwidth]{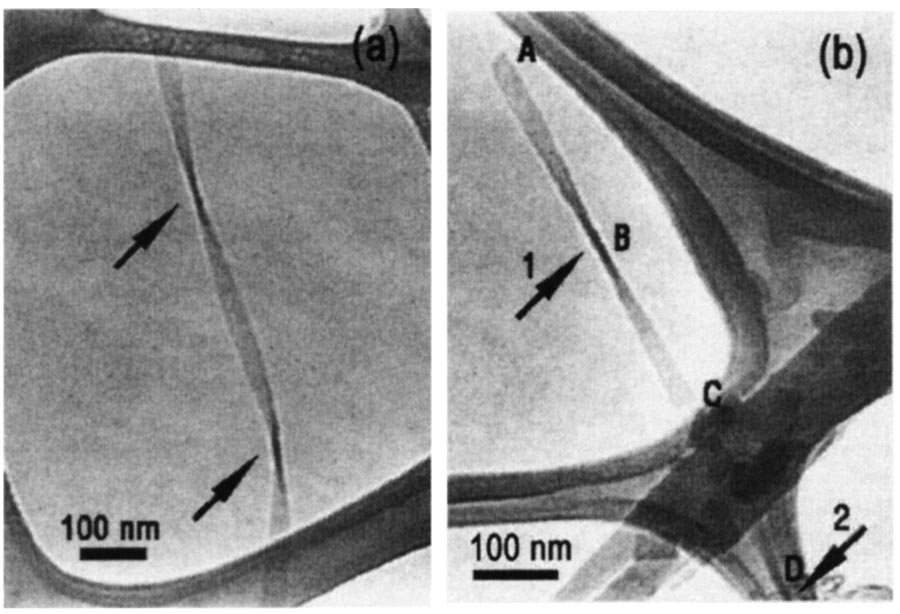}
\caption{TEM images showing a suspended, fully collapsed and twisted MWCNT (a), and a freestanding, fully collapsed and
twisted MWCNT (b), on a lacy carbon TEM grid \cite{yu.mf_20011, yu.mf_2004}.}
\label{fig.twisting}
\end{figure}

Upon compressing, bending, or twisting, structural deformations of CNT become more complicated. With an increasing
compression stress, different buckling patterns in CNT at the point of instabilities can be created
\cite{yakobson.bi_1996}. As shown in Figure \ref{fig.compressing}, each shape change corresponds to an abrupt release of
energy and a singularity in the stress-strain curve. MD simulations showed that the presence of intertube vdW
interactions tends to weaken the CNT bundles under compression \cite{liew.km_20061}. On the other hand, with increasing
the tube diameter, the bending modulus was found to decrease sharply \cite{poncharal.p_1999}, as a rippling mode becomes
energetically favorable \cite{lourie.o_1998, arroyo.m_2003}. The modeling on the wavelike distortion indicated that
there is a critical diameter at a given load and a CNT length, for the emergence of the rippling mode
\cite{liu.jz_2003}. It was also found that thick MWCNTs very prone to develop rippling deformations in bending and
twisting, governed by the interplay of strain energy relaxation and intertube interactions \cite{arroyo.m_2003,
qian.d_20031}. Figure \ref{fig.twisting} shows experimental observations partially collapsed segments in freestanding
and twisted MWCNTs with TEM \cite{yu.mf_20011}.

\section{Interlayer and intertube interactions}
\label{sec.interaction}

\subsection{Sliding between MWCNT shells}

Different from SWCNTs, the inter-layer interactions in MWCNTs exhibit interesting sliding phenomena during the tensile
stretching and self-oscillation. By modeling of the inter-layer interaction with different atomic potentials, such as
the classic Lennard-Jones (LJ) potential and the registry-dependent Kolmogorov-Crespi (KC) potential (zeroth generation
RDP0 \cite{kolmogorov.an_2000} and first generation RPD1 \cite{kolmogorov.an_2005}), could cause significantly different
energy corrugation between graphitic layers \cite{qian.d_2003, kolmogorov.an_2005, guo.wl_2005, zhang.xh_2013}.
Nevertheless, the inter-layer sliding has been systematically studied despite the type of atomic potential. Experiments
have also shown the inter-layer sliding and superlubricity of CNTs.

Experimental measurements on the dependence the sliding force against the contact length between the MWCNT shells
suggested several responsible sources including the surface tension, the shear elastic force, and the edge (tube
termination) effect force \cite{yu.mf_20002}. The shear strength was found to be 0.08--0.3 MPa depending on the
inter-layer commensurability. After the inner shells were pulled out from the outer shells, spontaneous retraction was
observed with TEM due to the attractive vdW forces between these shells \cite{cumings.j_2000}. The inter-layer static
and dynamic shear strengths were estimated to be 0.43--0.66 MPa. The telescoping motion of an MWCNT also showed an
ultralow friction, below $1.4\times10^{-15}$ N per atom ($8.7\times 10^{-4}$ \si{\milli\eV\per\angstrom})
\cite{kis.a_2006}. More direct observation of the superlubricity in double-walled CNTs (DWCNTs) was conducted by
extracting the inner shell out from a centimetres-long DWCNT, where the inner-shell friction (vdW force) was estimated
to be 1.37--1.64 nN (the total number of atoms was unknown, yet could be over $10^8$--$10^9$ for a tube length of 1 cm)
\cite{zhang.rf_20131}, corresponding to a friction force of about $10^{-6}$--$10^{-5}$ \si{\milli\eV\per\angstrom} per
atom. By following the low-friction induced telescoping, gigahertz oscillators were theoretically proposed based on the
oscillatory extrusion and retraction of the inner shells \cite{zheng.qs_2002, guo.wl_2003, legoas.sb_2003, zhao.y_2003,
rivera.jl_2003, liu.p_20052, zhao.xc_2006, li.b_20101}, and the inter-layer commensurability determines the rate of
energy dissipation during the oscillation \cite{guo.wl_2003}. The friction force between oscillatory layers was
estimated to range from $10^{-17}$ to $10^{-14}$ N per atom, about $6\times 10^{-6}$--0.006 \si{\milli\eV\per\angstrom}
\cite{zhao.y_2003}. Similarly, the rotation between CNT shells and the rotational friction were observed or measured by
experiments and simulations \cite{fennimore.am_2003, zhang.sl_2004, servantie.j_20061, cai.k_2015}. The rotational
friction is a broadband phenomenon, as it does not depend strongly on a specific vibrational mode, but rather appears to
occur from the aggregate interactions of many modes at different frequencies \cite{cook.eh_2013}.

\begin{figure}[t!]
\centering
\includegraphics[width=0.40\textwidth]{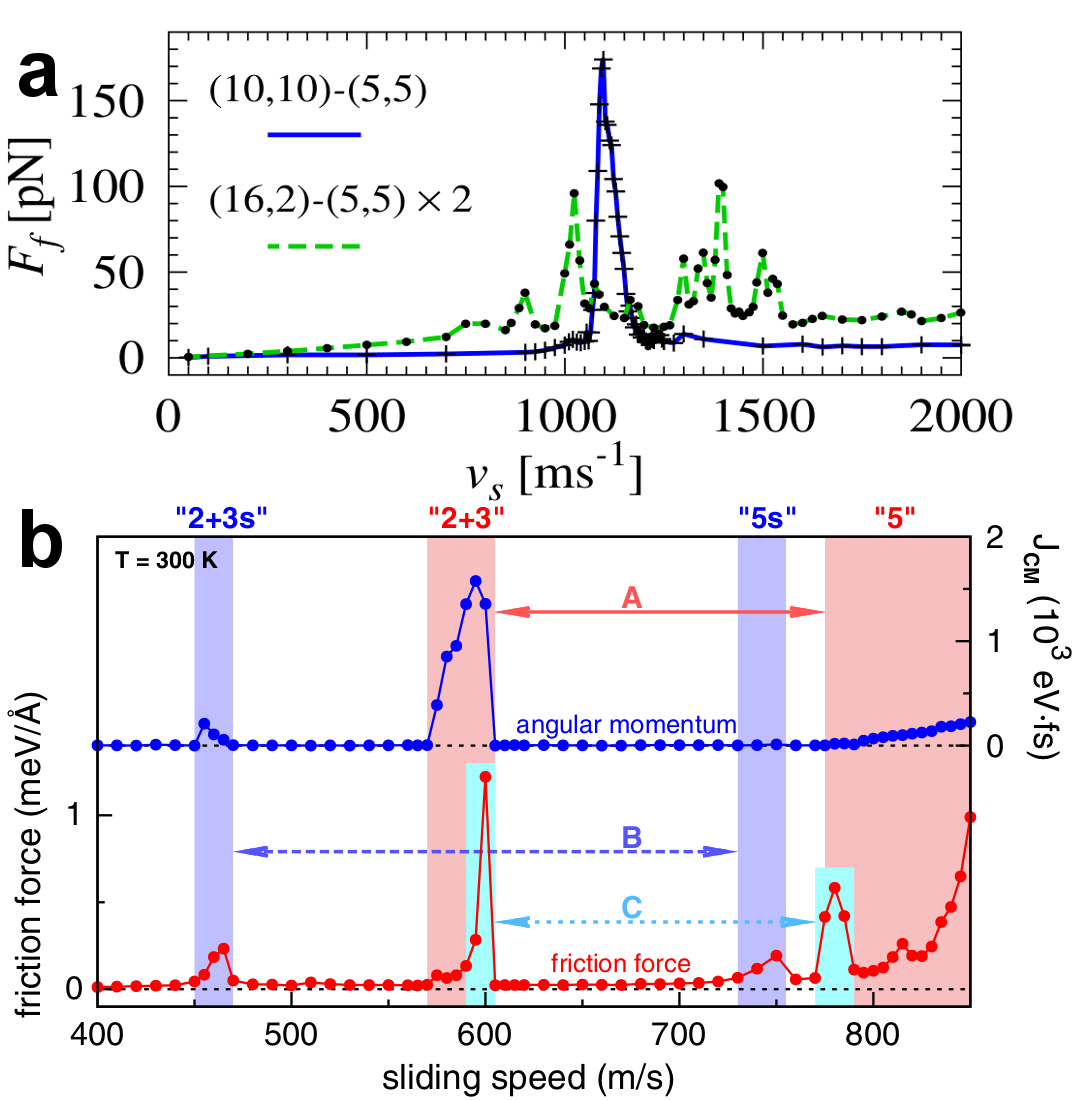}
\caption{(a) There is a massive increase in friction at $\sim$1100 \si{\m\per\s} for the (5,5)@(10,10) DWCNT
\cite{tangney.p_2006}. (b) The centre-of-mass angular momentum $J_\text{CM}$ (top) and the sliding frictional force per
inner tube atom (bottom) as functions of the sliding speed \cite{zhang.xh_2009}. A, B and C are three resonance regions
related the giant friction forces.}
\label{fig.friction}
\end{figure}

At high sliding speeds, the friction between CNT layers was not hydrodynamical, in fact not even monotonic with the
sliding speed \cite{tangney.p_2006}. A high speed can develop a sharp friction peak and onset parametric excitation of
CNT's ``breathing'' phonon modes. Figure \ref{fig.friction}a shows that there is a massive increase in sliding friction
at a speed of $\sim$1100 \si{\m\per\s} for a (5,5)@(10,10) DWCNT. The highest friction force could be up to 170 pN,
corresponding to $\sim$0.15 \si{\milli\eV\per\angstrom} per inner tube atom. The main source of the inter-layer friction
was believed to be the inner nanotube terminations \cite{tangney.p_2004}, while the ``bulk'' friction was unfortunately
considered to be irrelevant.

To understand the ``bulk'' frictional behavior, two groups of studies were carried out \cite{zhang.xh_2007,
zhang.xh_2009, zhang.xh_2013}: pulling simulations where a constant external force was applied on each atom of the inner
tube along the tube axis, and rigid sliding simulations where the inner tube was constrained to slide as a whole at a
fixed speed. The pulling simulation revealed a static friction (depinning) force of 0.02 \si{\milli\eV\per\angstrom} at
300 K and 0.1 \si{\milli\eV\per\angstrom} at 50 K for the (5,5)@(10,10) DWCNT. After the depinning, the speed of the
inner tube jumps directly to plateaus. Three speed plateaus are seen at $v\approx 450$, 720, and 780 \si{\m\per\s}. This
phenomenon is ubiquitous as it is also observed for an incommensurate case where a chiral (11,2) tube slides inside a
(12,12). The general occurrence of plateaus and jumps is a natural consequence of frictional peaks for growing the
speed. Figure \ref{fig.friction}b shows sharp frictional peaks near $v\approx 450$, 570, 720, and an important threshold
onset near 780 \si{\m\per\s}. These peaks are known to generally arise out of parametric excitation of the ``breathing''
phonon modes, classified by an angular momentum index $n$ (for tangential quantization around the tube axis). Such
nonmonotonic force-speed characteristics implies a ``negative differential friction'', whereby an increasing applied
force yields an inner tube sliding speed that grows by jumps and plateaus, rather than smoothly.

The surprise comes from analyzing the two parts of angular momentum $J = J_\text{CM} + J_\text{pseudo}$ around the tube
axis such as its center-of-mass (rigid body rotation with angular velocity $\omega$) and shape-rotation
(``pseudorotational'') parts \cite{zhang.xh_2009, zhang.xh_2013}. Generally, zero at generic sliding speed due to lack
of nanotube chirality, $J_\text{pseudo} = - J_\text{CM} = 0$, $J_\text{pseudo}$ jumps to nonzero values at the
frictional peaks and past the threshold, where $J_\text{pseudo} = - J_\text{CM} \neq 0$, see Figure \ref{fig.friction}b.
This is a spontaneous breaking of chiral symmetry occurring in nanoscale friction, whose mechanism is from the third-
and fourth-order energy nonlinearities. The inner-tube sliding causes a washboard frequency due to the lattice
periodicity $a_\text{CC} = 2.46$ \si{\angstrom}. At the critical frequencies, the `$n=5$' or `$2+3=5$' excitations are
observed, where the DWCNT's $n=5$ mode is resonantly excited or the $n=2$, 3, 5 modes are jointly excited by the simple
matching condition $2+3=5$. Besides these ``normal'' excitations (labelled A in Figure \ref{fig.friction}b), there are
`$n=5s$' and `$2+3s=5$' excitations, where the $3s$ and $5s$ are the single outer tube's phonon modes. These excitations
are labelled B in Figure \ref{fig.friction}b. The two C-peaks in Figure \ref{fig.friction}b are ascribed to the
phenomenon of strong stick-slip frictions, possibly related to the phonon modes along the tube axis.

\subsection{Intertube friction}

In most of the cases, individual CNTs are bundled together as a CNT rope, owing to the strong aggregation tendency
between adjacent CNTs. As mentioned above, experimental measurement showed a tensile strength just up to 13--52 GPa
\cite{yu.mf_20001}, far below the ideal strength for individual CNTs. The easy sliding between adjacent CNTs has become
a severe problem hindering their engineering applications. For example, slip rather than breakage of individual CNTs was
observed in the fracture test of polymers reinforced by CNT bundles \cite{ajayan.pm_2000}. As CNTs can contact with each
other in parallel way or at a certain cross angles, the interfacial shear strengths between CNTs should be discussed
separately.

The static and kinetic frictions between two perpendicular CNTs were investigated experimentally. A coefficient of
friction of 0.006$\pm$0.003 was obtained by sliding an MWCNT perpendicularly on an SWCNT surface, and the shear strength
was derived to be 4$\pm$1 MPa \cite{bhushan.b_2008}. The shear strength was one to two orders of magnitude larger than
the inter-shell shear strength of 0.05 MPa for MWCNTs in vacuum, possibly due to the presence of water at the intertube
interface in ambient. A following experiment, namely a vertical friction loop measurement, was proposed to characterize
the adhesion and friction properties between CNTs \cite{bhushan.b_20081}. An MWCNT tip was ramped in the vertical
direction against a suspended SWCNT. During ramping, a stick-slip motion was found to dominate the sliding, attributed
to the presence of defect-induced or amorphous-carbon-associated high energy points on the tip surface. Surprisingly,
the coefficients of static friction and shear strength were finally evaluated to be about 0.2 and 1.4 GPa, respectively.

MD simulations showed that the sliding behaviors and interfacial shear strengths between two crossly contacting CNTs are
much more complicated than the frictions between CNTs and graphite, graphite and graphite, and the inner and outer
shells of MWCNTs \cite{li.cx_2010}. The simulation was performed on two SWCNTs in contact at different cross angles. For
the parallel contact, the axial interfacial shear strengths between the zigzag-zigzag ($\sim$0.25 GPa) and
armchair-armchair pairs ($\sim$0.5 GPa) are two orders of magnitude larger than those of CNTs having different chirality
(0.5--1 MPa). CNTs with diameters larger than 1 nm will slide relatively from an AB stacking position to the next
nearest AB stacking position while smaller ones do not slide through AB stacking positions. For two cross contacting
CNTs, the magnitude of interfacial shear strength is much less dependent upon tube chirality. The highest value of shear
strength was reported to be about 1 GPa in SWCNT bundles, however, still as quite low as to be improved for applications
in high-performance composite materials \cite{salvetat.jp_1999}.

\subsection{Strategies to improve the intertube load transfer}

\begin{figure}[t!]
\centering
\includegraphics[width=0.48\textwidth]{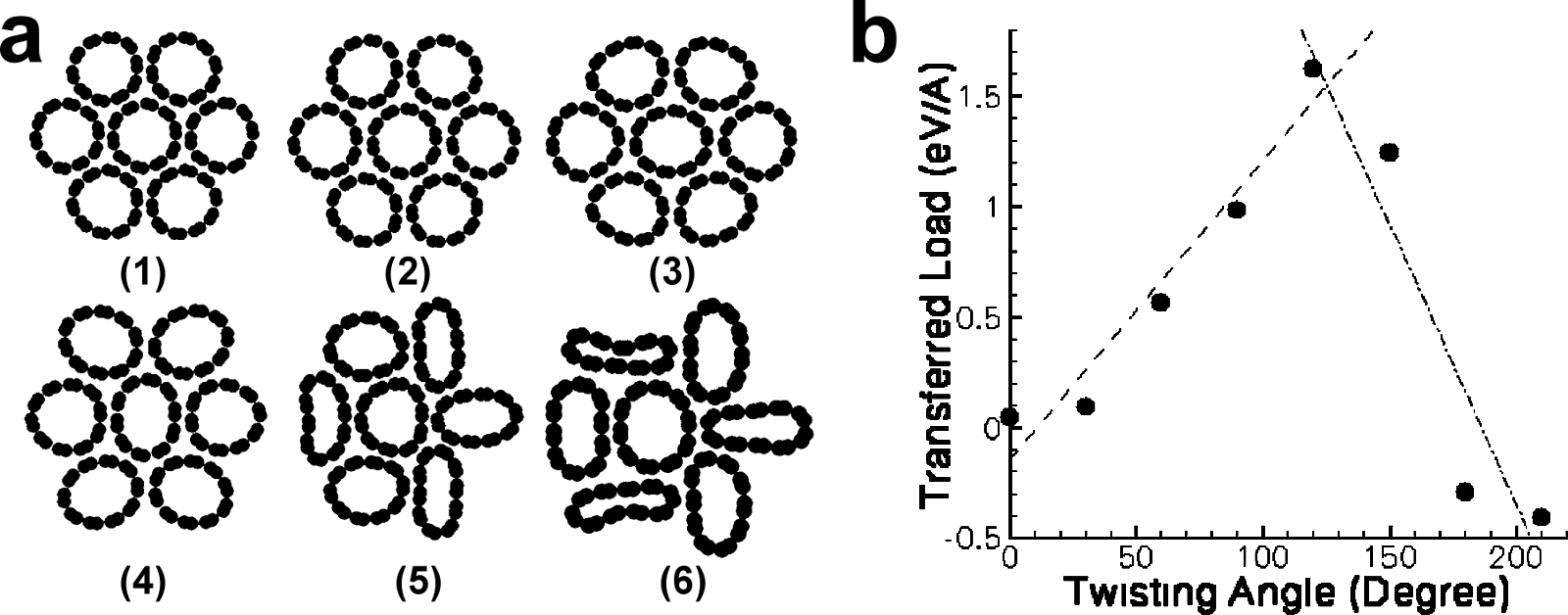}
\caption{(a) Change in cross section at the mid-length of a SWCNT bundle with increasing the twist level and (b) the
corresponding load transfer ability as a function of twist angle \cite{qian.d_2003}.}
\label{fig.deformation}
\end{figure}

A molecular mechanics modeling showed that in order to make the strength of a CNT bundle as high as that of the lowest
value of individual CNTs (11 GPa), the contact length should be up to $\sim$3800 nm, by assuming the CNTs are bundled
together without changing their circular cross sections \cite{qian.d_2003}. With considering the vdW-force-induced
radial deformation \cite{tersoff.j_1994}, the intertube distance-of-closest changes from 3.34 to 3.25 \si{\angstrom} for
(10,10) CNT bundles, the required contact distance for 11 GPa can decrease to $\sim$1300 nm. This implies that the key
to improve the intertube load transfer is the interface design between CNTs.

Twist is an effective way to induce strong radial deformation for CNT bundles. Figure \ref{fig.deformation}a shows the
cross sections obtained from MD simulations with increasing the twist level and Figure \ref{fig.deformation}b shows the
corresponding transferred load due to the shape deformation at twist angles of 0, 30, 60, 90, 120, 150, 180, and
210\si{\degree} \cite{qian.d_2003}.It was observed that the individual CNTs start to collapse in the cross section, and
therefore, more atoms are in close contact. For the case of no twist, a force of only $\sim$0.048 \si{\eV\per\angstrom}
is transferred to the center tube, while at 120\si{\degree} the transferred load increases to 1.63
\si{\eV\per\angstrom}. Clearly, the radial deformation strongly depends on the twist angle, which consequently changes
the nature of the contact and contributes to a new inter-layer tribology.

Even a slight structural deformation could cause an improved load transfer. For example, when a DWCNT was slightly bent
into a curved one, the energy transfer between the two tubes becomes much faster than that between two straight tubes
\cite{cai.k_2015}. Slight displacement of the inner-tube could add a large degree of disturbance to the DWCNT's original
state of equilibrium, and thus present strong inertia to oppose the extraction process, indicating the importance of the
inner tube as structural support for the outer tube \cite{kok.zkj_2016}.

\begin{figure}[t!]
\centering
\includegraphics[width=0.45\textwidth]{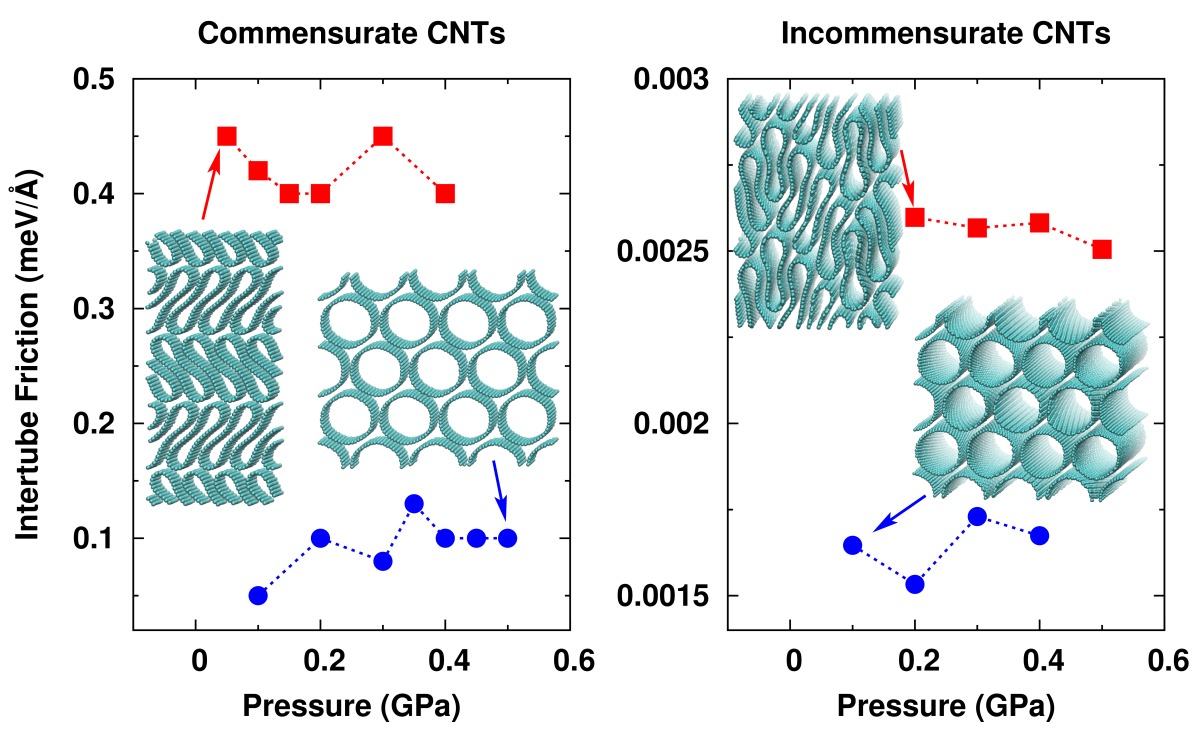}
\caption{The intertube friction force can be increased by a factor of 1.5--4, depending on tube chirality and radius,
when all tubes collapse above a critical pressure and when the bundle remains collapsed with unloading down to
atmospheric pressure \cite{zhang.xh_2010}.}
\label{fig.collapse}
\end{figure}

Pressure-induced structural phase transition could be a new strategy to improve the intertube load transfer
\cite{zhang.xh_2010}. For a CNT bundle containing 16 (23,0) SWCNTs, the intertube static friction can be improved from
0.1 \si{\milli\eV\per\angstrom} per atom to 0.4--0.45 \si{\milli\eV\per\angstrom} after all the tubes collapse under
pressure higher than 0.5 GPa, see Figure \ref{fig.collapse}. Due to the large tube diameter, the collapsed CNT structure
can be well maintained even by unloading the pressure down to 0.05 GPa. In order to realize a 10 GPa tensile strength,
the minimum tube length should be just 167 nm, much smaller than the length of 1090 nm before the phase transition. For
a CNT bundle containing totally different tube chirality, the dynamic friction force can be improved from 0.0017 to
0.0026 \si{\milli\eV\per\angstrom} after the phase transition. Such study also indicated that few-walled large-diameter
CNTs could be very important for engineering applications due to the easy ability to induce the collapse phase
transition.

\begin{figure}[t!]
\centering
\includegraphics[width=0.48\textwidth]{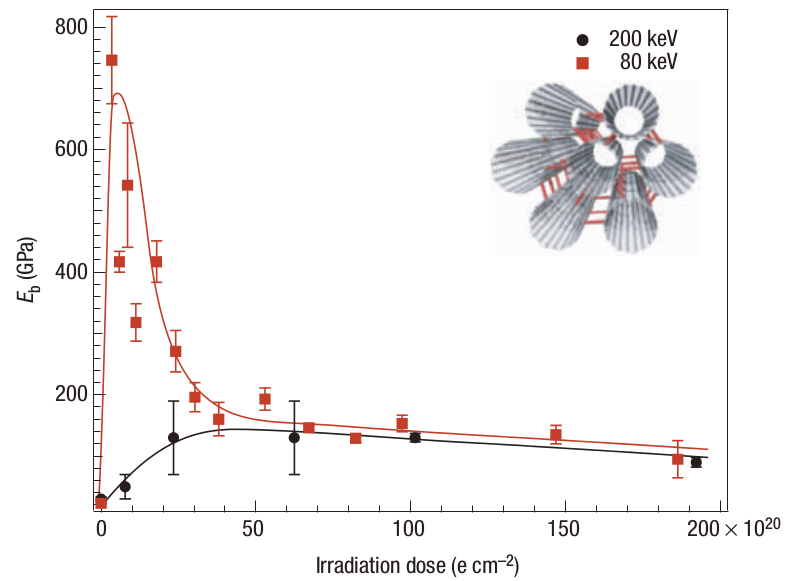}
\caption{Behaviour of bending modulus as a function of received dose for two incident electron energies
\cite{kis.a_2004}. For 200 keV, the modulus increases on short exposures, due to crosslinking and degrades at higher
exposures because of structural damage. The bundle irradiated with 80-keV electrons shows a much stronger and sharper
increase of the bending modulus.}
\label{fig.irradiation}
\end{figure}

Covalent bonding between the shells in MWCNTs or between adjacent CNTs can increase the inter-layer and intertube shear
strength by several orders of magnitude \cite{huhtala.m_2004, kis.a_2004, fonseca.af_2010}. The covalent inter-shell
bonds can be formed due to the on-shell vacancies or inter-shell interstitials. According to the types of defect, the
force needed to initiate sliding of the shells varies from 0.08--0.4 nN for a single vacancy to 3.8--7.8 nN for two
vacancies, an inter-shell interstitial, and an inter-shell dimer \cite{huhtala.m_2004}. Therefore, small-dose electron
or ion irradiations are suggested to partially transfer the load to the nanotube inner shells. MD simulations revealed
that inter-wall sp$^3$ bonds and interstitial carbon atoms can increase load transfer between DWCNT walls and that
inter-wall sp$^3$ bonds are the most effective \cite{fonseca.af_2010}. Similarly, by using moderate electron-beam
irradiation inside a TEM, stable links between adjacent CNTs within bundles were covalently formed \cite{kis.a_2004}. At
a high irradiation energy of 200 keV, there was a substantial increase in effective bending modulus and thus an increase
in shear modulus for low doses, while a long irradiation time rather damaged the well-ordered structure of tubes (Figure
\ref{fig.irradiation}). At 80 keV, electrons were shown not to damage isolated CNTs but had enough energy to lead to the
formation of mobile interstitial atoms in the confined space between adjacent CNTs. As a result of the newly formed
covalent bonds, a huge increase of the bending modulus by a factor of thirty were observed (Figure
\ref{fig.irradiation}). However, a long-time dose at low energies could still cause the accumulation of damage to the
structure, making the mechanical performance decrease eventually.

Even without introducing covalent bonding between the sp$^2$ layers, the existence of structural defects in either CNT
layer could lead to a sharp increase in friction and energy dissipation rate \cite{guo.wl_2005}. 

However, besides the tube structure, defect level, and intertube contact, there are many other issues in macroscopic CNT
assembly materials. For example, the CNT packing density, alignment, entanglement, aggregation size, and surface
functionalization could be important structural parameters, as to be discussed below.

\section{Carbon nanotube assembly materials}
\label{sec.assembly}

So far, CNT fibers, films, forests, and gels are well-known and typical macroscopic assembly materials where the CNTs
are the main constituents, especially at a mass fraction much more than 50\% \cite{liu.lq_2011}. For example, CNT fiber
is a one-dimensional assembly containing millions of individual tubes \cite{behabtu.n_2008, zhang.xh_2012, lu.wb_2012,
miao.mh_2013, di.jt_20161}. It has been found that CNT fibers could have much higher specific modulus and specific
strength than those of commercial carbon and polymeric fibers. For the composite materials, there have been two
different applications of CNTs such as a dispersion-based fabrication by using CNTs as reinforcing fillers in a polymer
matrix \cite{coleman.jn_2006, moniruzzaman.m_2006} and an assembly-based fabrication by directly assembling CNTs into
composites \cite{zhang.xh_2016}. In the former, the CNT content is strongly limited to be no larger than $\sim$5 wt\%,
while in the latter, a small amount of polymer molecules (about 20--30 wt\%) are used to strengthen the interfaces
between CNTs. To describe these macroscopic CNT materials, different structural parameters are proposed and investigated
in the past decade.

\begin{figure*}[t!]
\centering
\includegraphics[width=0.70\textwidth]{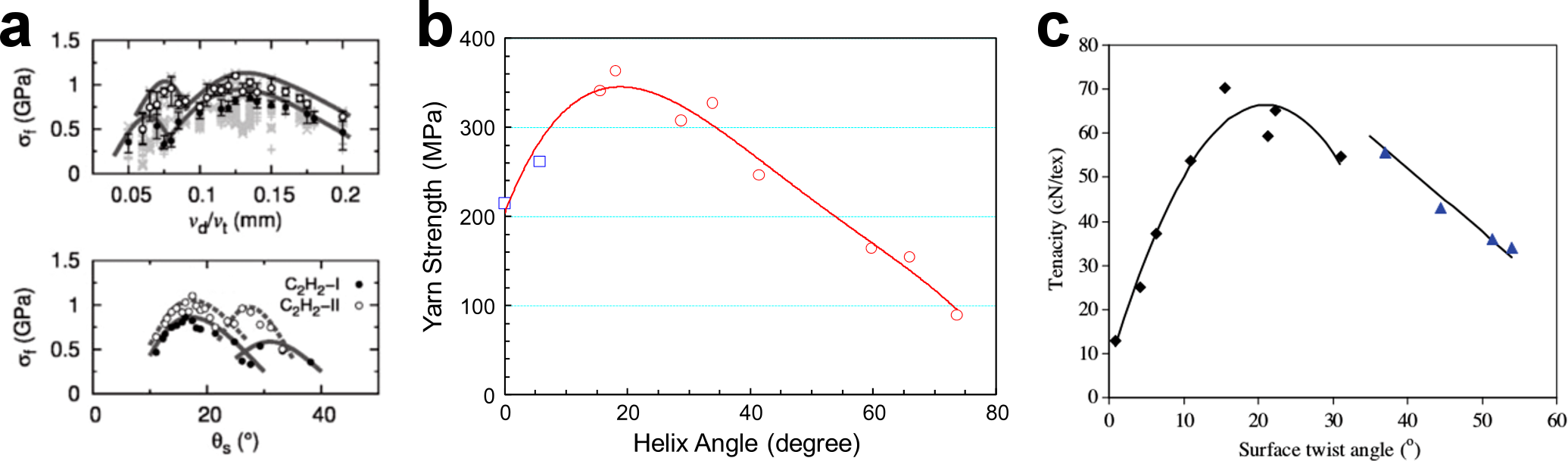}
\caption{(a) Double-peak strength behavior of CNT fibers with increasing twist angles \cite{zhao.jn_2010}. The surface
twist angle ($\theta_s$) depends on the twisting and drawing speeds ($v_t$ and $v_d$) by $\theta_s = \arccos
[(v_d/v_t)/\sqrt{(\pi d_\text{f})^2 + (v_d/v_t)^2}]$, where $d_\text{f}$ is the fiber diameter and was $>$10.5
\si{\micro\m}. (b) Dependence of yarn tensile strength on helix angle for forest-spun MWCNT fibers \cite{fang.sl_2010}.
The fiber diameter was $\sim$20 \si{\micro\m}. (c) Relationship between the fiber's specific strength or tenacity with
the surface twist angle \cite{miao.mh_2010}.}
\label{fig.twist}
\end{figure*}

\begin{figure*}[t!]
\centering
\includegraphics[width=0.70\textwidth]{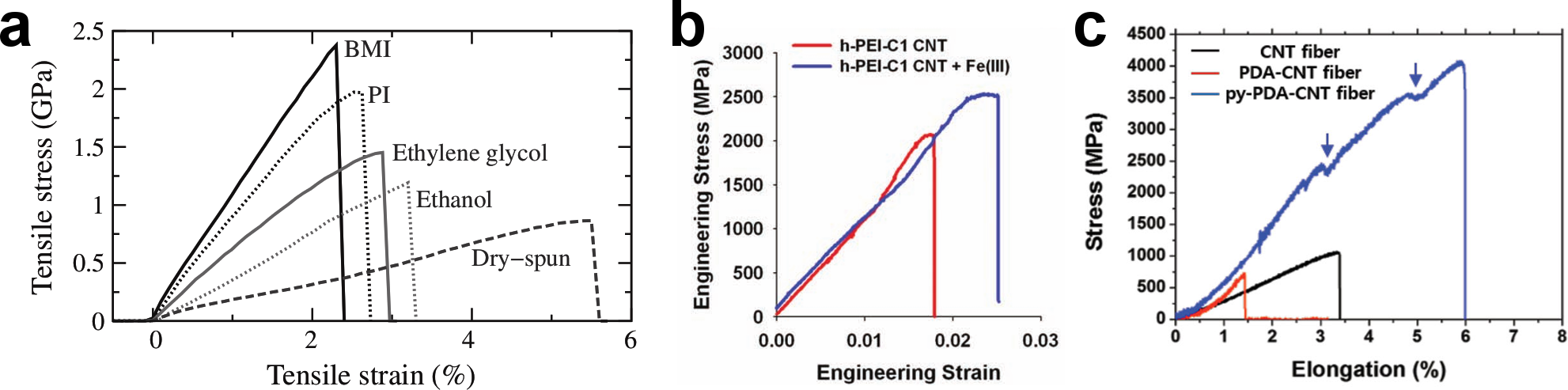}
\caption{(a) Stress-strain curves of dry-spun, ethanol-densified, ethylene-glycol-densified, PI-reinforced, and
BMI-reinforced CNT fibers \cite{li.s_2012}. (b) Stress-strain curves of PEI-catechol-reinforced CNT fibers, including
one after addition of Fe(III) \cite{ryu.s_2011}. (c) Stress-strain curves of as-drawn CNT fiber, CNT/PDA fiber, and
CNT/(pyrolyzed PDA) fiber \cite{ryu.s_2015}.}
\label{fig.fiberstrength}
\end{figure*}

\subsection{CNT fibers}

CNT fiber can be produced by the coagulation-based ``wet spinning'' \cite{vigolo.b_2000, dalton.ab_2003,
ericson.lm_2004, behabtu.n_2013}, the ``direct spinning'' from a CNT aerogel \cite{zhu.hw_2002, li.yl_2004,
koziol.k_2007} or similarly from a pre-formed CNT film \cite{ma.wj_2009}, and the ``forest spinning'' based on
vertically aligned CNT forests \cite{jiang.kl_2002, zhang.m_2004, li.qw_2006}. A CNT fiber with a diameter of $\sim$10
\si{\micro\m} usually contains more than $10^6$ individual CNTs. Such assembly feature is different from those
well-known fibers for many years, which are usually a solid structure without internal interfaces. CNT fiber can also be
considered a continuous length of interlocked ``filaments'' (CNT bundles), where the bundles are formed during the CNT
growth rather than in the spinning process. As I have pointed out recently \cite{zhao.jn_2015}, although also being
called as CNT yarn, CNT fiber is indeed not a yarn because the CNT bundles are not macroscopically processable. On the
contrary, the basic components of a yarn, the long and parallel or interlocked filaments, are usually processable
objects with a width larger than several micrometers \cite{hearle.jws_20011}.

Twist is the most fundamental treatment to make the CNTs assembled into a continuous fiber with a circular cross
section. The as-spun fiber is relatively loose with noticeable spaces between CNTs or CNT bundles when the twist angle
is small. Increasing the twist angle is an effective method to densify CNT fibers, and thus improves the friction
coefficient between the CNTs. Figure \ref{fig.twist} shows  the strength-twist relationship for different CNT fibers
\cite{zhao.jn_2010, fang.sl_2010, miao.mh_2010}. All studies showed that the highest fiber strength was found at a twist
angle of 15--20\si{\degree}. However, with further increasing the twist, the load (upon stretching) compresses the fiber
rather than acts totally along the fiber axis due to the twist angle between the CNTs and the axis, making the
utilization efficiency decreases at high twist angles. The experimental observations that the strength initially
increases with twist angle and then decreases are almost the same as those in traditional textile yarns. Nevertheless,
the constituent CNTs are different from the filaments in a textile yarn;  the CNTs are hollow cylinders and can deform
the cross section under pressure due to a structural phase transition \cite{elliott.ja_2004, zhang.xh_2004,
zhang.xh_20041, sun.dy_2004, ye.x_2005, gadagkar.v_2006}. For those fibers spun from few-walled CNTs (wall numbers
smaller than 6), it can be seen that beyond the optimal twist angle, an additional strength peak is at large twist angle
of 27--30\si{\degree}. This second peak arises from the collapse transformation of the CNT hollow structure, which
renders the CNT bundles much stronger and reduces the bundle cross-sectional area.

Solvent densification is another interfacial treatment to CNT assemblies, due to a large capillary force
\cite{liu.h_2004, chakrapani.n_2004}. Therefore, liquid infiltration by using water, ethanol, acetone or dimethyl
sulphoxide (DMSO) is also often used in the method of array spinning to densify CNT fibers \cite{zhang.xb_2006,
liu.k_2010, li.s_2012, qiu.j_2013}. Although the densification process does not improve nanotube orientation, it
enhances the load transfer between the nanotubes, thus ensuring that most of them are fully load-bearing. For example,
the fiber diameter was shrunk from 11.5 to 9.7 \si{\micro\m} after acetone shrinking \cite{liu.k_2010}. However, the
role of solvent is hardly known as the capillary force can be influenced by the solvent’s volatility (boiling point),
surface tension, and interaction with CNT surfaces. By comparing various solvents including the nonpolar solvents of
$n$-hexane, cyclohexane, cyclohexene, toluene, and styrene, polar protic solvents of glycerine, methanol, ethanol,
water, ethylene glycol, and 1,3-propanediol, and polar aprotic solvents of ethyl acetate, acetone, acetonitrile,
$N$,$N$-dimethylformamide (DMF), DMSO, and $N$-methyl-2-pyrrolidone (NMP), only few highly polar solvents show high
densifying ability to CNTs, namely DMF, DMSO, NMP and ethylene glycol, despite of their high boiling points and low
evaporation rates \cite{li.s_2012}. These solvents increased the fiber strength from 864 MPa (dry-spun) or 1.19 GPa
(ethanol) to 1.14--1.35 GPa (DMF, DMSO, NMP) or 1.33--1.58 GPa (ethylene glycol). Clearly, the solvent polarity shows a
key role in determining the capillary, as the polarity induced attractive binding energy with CNT ($E_\text{ind}$) goes
up quadratically with the solvent's local dipole moment $\mu$, by $E_\text{ind} = - \mu^2 \alpha / (4 \pi  \epsilon_0)^2
r^6$, where $\alpha$ is the static polarizability of CNT,  $\epsilon_0$ is the vacuum permittivity, and $r$ is the
distance between the dipole moment and the CNT surface \cite{inhetpanhuis.m_2003, israelachvili.jn_1991}.

\begin{figure}[t!]
\centering
\includegraphics[width=0.45\textwidth]{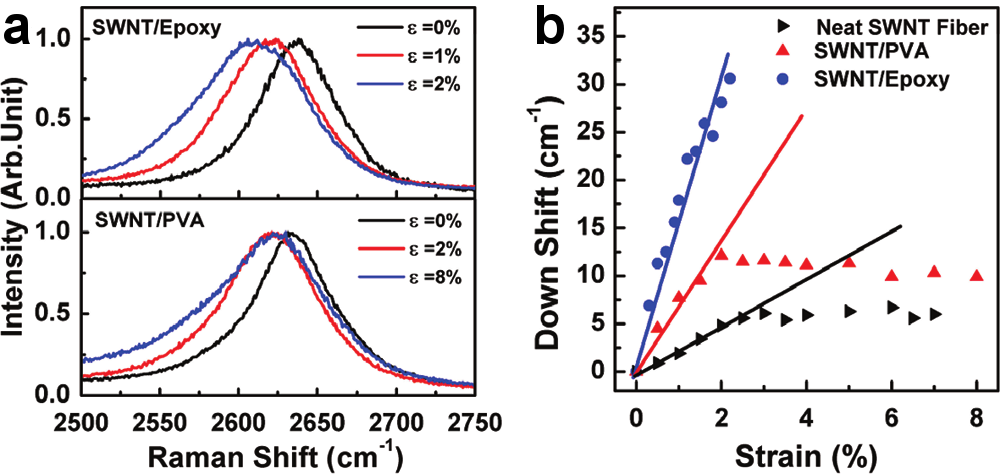}
\caption{Variations of Raman spectra under strain in different systems \cite{ma.wj_2009}. (a) Typical G' band Raman
spectra of epoxy- and PVA-infiltrated CNT fibers. (b) Downshifts of the peak position in different systems.}
\label{fig.raman}
\end{figure}

In addition to twisting and liquid densification, polymer impregnation is another effective treatment to enhance the
mechanical properties of CNT fibers. The polymers are introduced to bridge nonneighboring CNTs, such as
polyethyleneimine (PEI) \cite{ryu.s_2011}, polydopamine (PDA) \cite{ryu.s_2015} polyvinyl alcohol (PVA)
\cite{ma.wj_2009, liu.k_20101, li.s_2012}, polyacrylate \cite{naraghi.m_2010}, polyvinylidene fluoride (PVDF)
\cite{zou.jy_2016}, epoxy \cite{ma.wj_2009}, polyimide (PI) \cite{fang.c_2010, li.s_2012}, and bismaleimide (BMI)
\cite{li.s_2012, meng.fc_2014}. Figure \ref{fig.fiberstrength} shows several typical stress-strain curves for
polymer-reinforced CNT fibers, such as PI, BMI, PEI, and PDA \cite{li.s_2012, ryu.s_2011, ryu.s_2015}. The enhanced
mechanical properties of CNT fibers with polymer impregnation are attributed to couplings between the CNT network and
polymer chains occurring at the molecular level. As a result, the sliding between CNTs can be remarkably hindered by the
polymer network, especially that formed by the cured thermosetting polymers. Therefore, more external strain can be
transferred to the CNTs as shown in Figure \ref{fig.raman}, where the Raman peaks of CNTs down-shift to a lower
wavenumber by different values under axial strains \cite{ma.wj_2009}. Both PVA and epoxy enhanced the down-shifts,
corresponding to the improved load transfer between CNTs. The cured thermosetting polymer network also shows advantages
in improving the fiber modulus, as indicated by the largest change in G'-peak wavelength. Similarly, PI and BMI are
other efficient thermosetting polymers for producing high-modulus CNT fibers \cite{fang.c_2010, li.s_2012,
meng.fc_2014}. However, other polymers might be good at improving the fiber's strechability and toughness by preventing
the sliding-induced fiber fracture, possibly due to the polymer wrapping around the CNT connections \cite{zou.jy_2016},
cooperative deformation mechanisms of the soft and hard segments \cite{naraghi.m_2010}, iron-mediated covalent
cross-linking \cite{ryu.s_2011}, or cross-linked polymeric binders \cite{ryu.s_2015}.

The structure of CNT fiber is hierarchical, with CNTs self-assembling into bundles and the bundles forming an aligned
network in the fiber \cite{vilatela.jj_2011, deng.wl_2014}. Other assembly parameters also include the tube waviness or
straightness \cite{jia.jj_2011}, which can also result in different elasticity and modulus for CNT fibers. Notice that,
the above mechanical improvements are irrelevant to the micro structure of individual CNTs but are relevant to the
macroscopic assembly parameters. From the point of view of tube structure, few-walled CNTs have shown advantages in
producing high-performance fibers, rather than the ideal SWCNTs \cite{jia.jj_2011}. For the SWCNTs, usually 1--1.5 nm in
diameter, simulations and experiments have shown that a certain high pressure (hundreds of MPa) is necessary to
introduce even a slight deformation of their cross sections. That means, the intertube contact area is far limited
compared with the overall circumference of the cross section. While for the few-walled tubes, the tube diameter can be
large up to $\geq$5 nm, making it much easy to deform the cross section and increase the contact area. Experiments even
have shown, upon certain twisting and/or tensioning, collapse of few-walled tubes were possibly introduced
\cite{zhao.jn_2010, motta.m_2007}.

Furthermore, covalent interface engineering could an efficient post-treatment for the as-produced CNT fibers. Acid
treatment could introduce rich hydroxyl, methyl/methylene, and carbonyl groups onto CNT surfaces, which also made CNT
fibers shrunk slightly, or more densified. Such treatment enhanced the load transfer between CNTs, according to a
theoretical study \cite{kis.a_2004}, and thus resulted in an increase in fiber strength and modulus
\cite{meng.fc_20121}. Recently, by using an incandescent tension annealing process (ITAP), where the current-induced
Joule heating increase the temperature of CNT fiber up to 2000 \si{\celsius} in vacuum, sp$^3$ covalent bonding was
observed in CNT fibers \cite{di.jt_2016}. As a result, the fiber modulus could be improved from 37 to 170 GPa, once the
ITAP was performed with maintaining a high tension on the fiber.

In order to further improve the utilization efficiency of the individual CNT's mechanical properties in CNT fibers, more
effort should be paid to the delicate combining of the choice of CNT structure, controlling in CNT packing density,
alignment, and straightness, and intertube mechanics engineering by polymers or covalent functionalization.

\subsection{CNT films}

CNT films are another important macroscopic material with high mechanical performances. Various processing methods have
been developed to produce CNT films by using pre-formed CNT assemblies, among which these two methods are of great
importance: the layer-by-layer stacking of aligned CNT sheets and the stretching on entangled CNT webs
\cite{zhang.xh_2016}. Despite of the fabrication methods, there are also many common problems concerning with the
interface mechanics, as the CNT fibers do.

\begin{figure}[t!]
\centering
\includegraphics[width=0.48\textwidth]{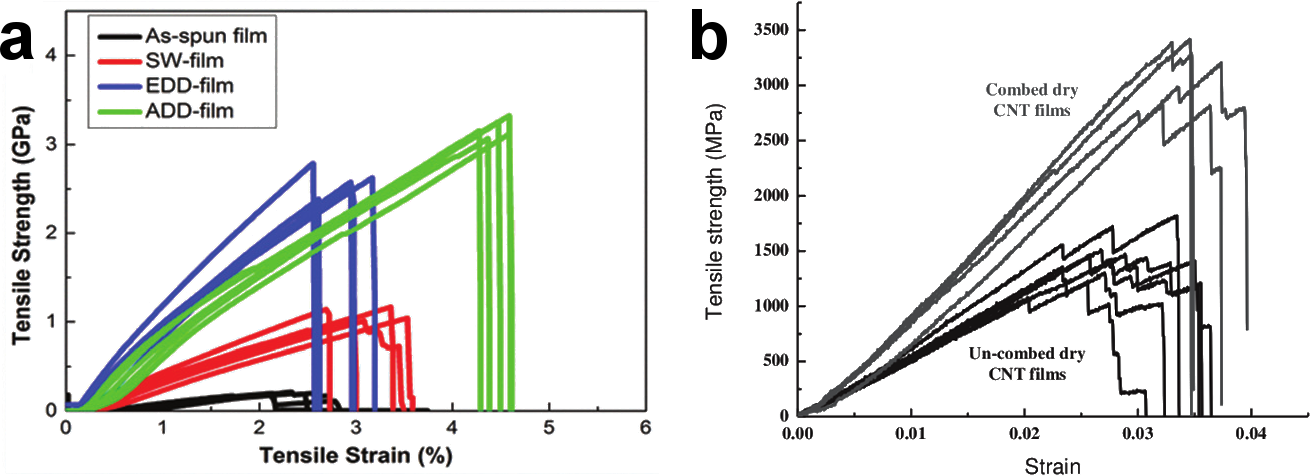}
\caption{Stress-strain curves for CNT films obtained by a stretch-dip-drying method (a) \cite{wang.yj_2015} and a
micro-combing method (b) \cite{zhang.lw_2015}. The films treated with acetone (ADD) and ethanol (EDD) were much stronger
than the as-spun and spray-wound (SW) films. The films after micro-combing had a larger Young's modulus than the ADD and
EDD films.}
\label{fig.aligned}
\end{figure}

For instance, liquid densification is widely used to CNT films \cite{di.jt_2012, wang.yj_2015, zhang.lw_2015}, and
polymer impregnation can remarkably improve the interfacial load transfer \cite{liu.w_2011, wang.x_20131, cheng.qf_2009,
cheng.qf_2010, jiang.q_2014, han.y_2015}. Different from CNT fibers where the tubes are assembled with the aid of
twisting, there is more freedom to tune the assembly structure. The CNTs can be got super-aligned by stretching the
films in a wet environment \cite{wang.yj_2015}, or by micro-combing before they are layer-by-layer stacked
\cite{zhang.lw_2015}. After improving the CNT alignment, the pure CNT films could be as strong as 3.2 GPa in tensile
strength, with a Young's modulus of 124--172 GPa \cite{wang.yj_2015,zhang.lw_2015}. Figure \ref{fig.aligned} shows the
stress-strain curves for the CNT films being both densified and super-aligned. Once these aligned CNTs were impregnated
with polymers, especially thermosetting polymers, the film's strength and modulus could be further improved. At a mass
fraction of 50--55\% for CNT, the CNT/BMI composite films exhibited a strength of 3.8 GPa and a modulus of 293 GPa
\cite{wang.x_20131}. Such superior properties are derived from the long length, high mass fraction, good alignment and
reduced waviness of the CNTs.

The stretching on entangled CNT webs is another way to obtain highly aligned CNT films. With the aid of BMI impregnation
and curing, the strength and modulus of the CNT/BMI composite films could be improved from 620 MPa and 47 GPa
(un-stretched) to 1600 MPa and 122 GPa, 1800 MPa and 150 GPa, and 2088 MPa and 169 GPa after being stretched by 30, 35,
and 40\% respectively \cite{cheng.qf_2009}. By further introducing surface functionalization for CNTs, the strength and
modulus were surprisingly improved up to 3081 MPa and 350 GPa, respectively \cite{cheng.qf_2010}.

\begin{figure}[t!]
\centering
\includegraphics[width=0.48\textwidth]{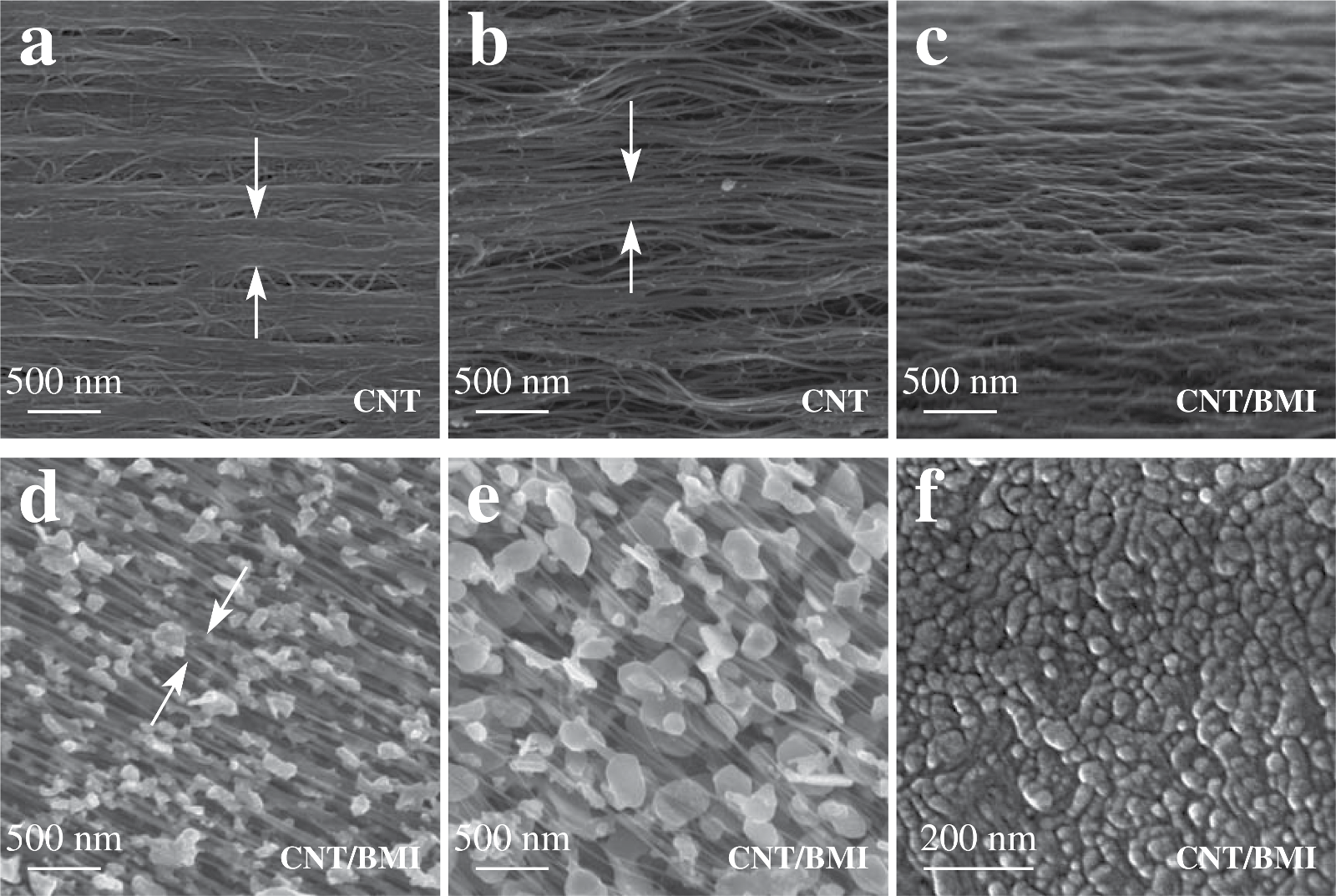}
\caption{Comparison of CNT assembly structure for different films \cite{han.y_2015}. (a,b) CNT aggregation in the
layer-by-layer stacking of array-drawn CNT sheets and the stretched dry films composed by entangled CNTs, respectively.
(c) The small-sized CNT bundles did not aggregate but were surrounded by BMI polymers and uniformly distributed. (d,e)
750 \si{\celsius} treated CNT/BMI composite films where the stretching was performed before and after resin
impregnation, respectively, showing the aggregation and unaggregation. (f) Cross section of the optimal CNT/BMI
composite structure by using focused ion beam treatment, indicating an aggregation size of 20--50 nm for the CNTs.}
\label{fig.aggregation}
\end{figure}

However, the direct stacking of CNT sheets and stretching on CNT webs could not avoid the CNT aggregation due to the vdW
interactions. As shown in Figure \ref{fig.aggregation}a and b, the aggregation size was up to hundreds of nanometer
\cite{han.y_2015}. Such aggregation of nanometer-sized components has become a severe problem in paving the way to
stronger materials \cite{wagner.hd_2007}. In the aggregation phase, the intertube load transfer is not as efficient as
at the CNT-polymer interface. Thus, such aggregation phase becomes the weak parts in the composites and hinders further
reinforcement. In an ideal structure, the nanometer-sized components should be uniformly distributed in the matrix
without forming any aggregation phases. Therefore, all the interfaces can play roles in shear load transfer. As inspired
by the formation process of natural composite structures, the BMI resins were infiltrated into CNT webs before any
stretching was performed. Since the liquid treatment just densifies the CNT network without inducing CNT aggregation,
the impregnation before stretching causes each CNT or CNT bundle uniformly covered by a thin layer of polymers.
Therefore, the following stretching could just align the CNTs but avoid the aggregation due to the thin layer of
polymer, resulting in an optimized composite structure. By adapting a multistep stretching process, the final composite
film exhibited superior mechanical performances. The highest tensile strength and modulus were up to 5.77--6.94 GPa and
212--351 GPa at a CNT-to-resin mass ratio of 7:3 and a stretching magnitude of 34\% \cite{han.y_2015}.

Figure \ref{fig.aggregation}c shows the morphology of the optimized CNT/BMI composites, where the CNT bundles were found
to be uniformly distributed and aligned. To characterize the aggregation of CNTs, thermal treatment and focused ion beam
treatment were used. After being heated to 750 \si{\celsius} to decompose the polymers, the remaining CNTs were found to
be aggregated/unaggregated if the stretching was performed before/after the polymer impregnation, see Figure
\ref{fig.aggregation}d and e. The direct observation of the cross section with focused ion beam treatment showed an
aggregation size of just 20--50 nm (Figure \ref{fig.aggregation}f), a size for the bundling that was still difficult to
avoid during the growth process.

On the whole, for the macroscopic CNT assemblies, the interface mechanics becomes much more complicated as the material
exhibits a collective dynamics from the constituent CNTs. Assembly parameters, including CNT packing density, alignment
and/or entanglement, twisting level, cross-linking, and aggregation size, are found to be critical for the mechanical
properties. By controlling the interface mechanics in the CNT assemblies, we can not only optimize the mechanical
properties, but also introduce multifunctionalities, as discussed below.

\subsection{Multifunctionalization by interface engineering}

\begin{figure}[t!]
\centering
\includegraphics[width=0.48\textwidth]{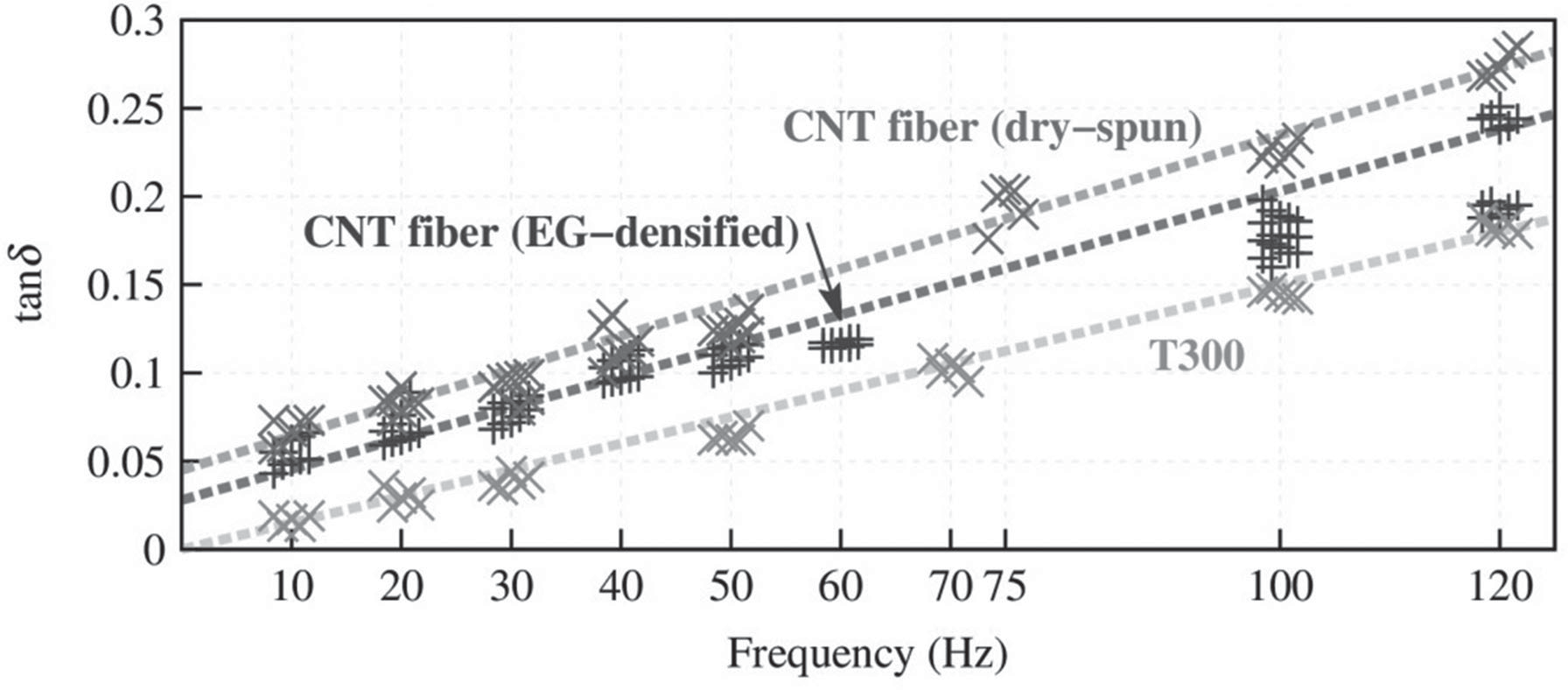}
\caption{The frequency dependence of loss tangent for the dry-spun CNT fiber, ethylene-glycol-densified CNT fiber, and
T300 carbon fiber \cite{zhao.jn_2015}.}
\label{fig.loss}
\end{figure}

Besides load transfer at the intertube interfaces, energy dissipation is also a common interfacial phenomenon. As CNT
fiber is formed by assembling millions of individual tubes, the assembly features provide the fiber with rich interface
structures and various ways of energy dissipation, including the internal viscosity and intertube friction
\cite{zhao.jn_2015}. Therefore, a modified Kelvin-Voigt model was adopted, where an elastic spring $K$, a viscous
damping coefficient $\eta$, and an intertube friction $f$ are connected in parallel \cite{murayama.t_1979}. Due to the
friction, there exists a friction-dependent component in loss tangent, which is not dependent on the frequency. As shown
in Figure \ref{fig.loss}, the loss tangent was nonzero (about 0.045) for the dry-spun CNT fibers, while that for the
T300 carbon fiber was nearly zero due to the absence of interfaces. When the CNT fibers were densified by ethylene
glycol, the CNTs became constrained, corresponding to reduced loss tangents, see Figure \ref{fig.loss}. Based on the
friction-based Kelvin-Voigt model, the damping performance of CNT fibers can be tuned in a very wide range. The
introduction of thermosetting polymers further decreases the loss tangent, while plying CNT fibers into a yarn increases
the loss tangent \cite{zhao.jn_2015}.

The CNT entanglement provides another way to enhance the damping ability \cite{liu.ql_2015}. The as-produced CNT films
based on an injection chemical vapor deposition exhibited a loss tangent of 0.37--0.42 at 200 Hz \cite{liu.ql_2015}, or
0.2 at 50 Hz \cite{han.y_2015}. After the network was stretched and strengthened with BMI, the loss tangent could
decrease remarkably down to about 0.05 at 50 Hz \cite{han.y_2015}.

When an electric current is passing through CNT assemblies, there exist parallel currents along the aligned CNTs. The
collective electromagnetic force between parallel currents can cause a volume contraction for the assembly
\cite{guo.wh_20121}. Based on such mechanism, CNT fibers can be developed as electro-mechanical actuator. Detailed study
revealed that the electro-mechanical coupling is quite complicated beyond the actuation \cite{meng.fc_2014}. For
example, when a current passes through a CNT fiber, the fiber's modulus could decrease remarkably due to the
current-induced weakening in \ce{C-C} $\sigma$ bond. Besides the electro-mechanical coupling, the current could induce a
strong Joule heating, making the fiber temperature increase very sharply. Such electro-thermal coupling can be used for
fast curing to the impregnated thermosetting resins at a moderate temperature of 200--300 \si{\celsius}
\cite{meng.fc_2014}, or for the formation of intertube sp$^3$ covalent bonding at a high temperature of 2000
\si{\celsius} in vacuum \cite{di.jt_2016}.

\section{Conclusion}

The control of interface mechanics is a key strategy to develop high-performance CNT assembly structures. High tensile
properties (strength up to 3--6 GPa and modulus up to 200--350 GPa) can be obtained for aligned CNT fibers and films,
corresponding to an improved utilization efficiency of CNT's mechanical properties. Furthermore, the dynamical
performance and multifunctionalities both depend strongly on the interface between CNTs, also in terms of CNT packing
density, alignment, entanglement, twisting level, cross-linking, and aggregation size. For future development of
high-performance CNT assembly materials, more delicate design on the multi-scale interfaces is required. I hope this
review could cast lights for the interface design ranging widely from the nanometer scale to the macroscopic scale. 

\begin{acknowledgments}
The author thanks Prof.\ Xingao Gong of Fudan University, Prof.\ Erio Tosatti of the International School for Advanced
Studies, and Prof.\ Qingwen Li of Suzhou Institute of Nano-Tech and Nano-Bionics for their kind collaborations, which
are the important sources for this chapter. Financial supports from the National Natural Science Foundation of China
(11302241, 51561145008), Youth Innovation Promotion Association of the Chinese Academy of Sciences (2015256), and
National Key Research and Development Program of China (2016YFA0203301) are acknowledged.
\end{acknowledgments}

%\bibliographystyle{nunsrt}
%\bibliography{sample.bib}

\end{document}